\documentclass{elsart}
\input epsf
\usepackage{graphicx}
\begin{document}
\runauthor{Thun and McKee}
\journal{Physics Letters B}
\begin{flushright}
{\bf UM HE 98-11}
\medskip
\end{flushright}
\begin{frontmatter}
\title{Interpreting the Atmospheric Neutrino Anomaly}

\author[Paestum]{R.P. Thun and S. McKee }

\address[Paestum]{Department of Physics, University of Michigan, Ann Arbor, MI
48109 USA, E-mail: rthun@umich.edu}

\begin{abstract}
We suggest that the atmospheric neutrino anomaly observed in the
Super-Kamiokande (and other) experiments results from the combined effects of
muon-neutrino to tau-neutrino oscillations with a $\Delta m^2$ value of
approximately 0.4 $eV^2$ and oscillations between muon neutrinos and electron
neutrinos (and vice-versa) with 0.0001$<\Delta m^2 <0.001~eV^2$.  With an
appropriate choice of a three-neutrino mixing matrix, such a hypothesis is
consistent with essentially all neutrino observations.
\end{abstract}

\end{frontmatter}
\pagebreak

\section{Introduction}
Statistically significant anomalies have been reported in observations of solar
neutrinos \cite{allison,fukuda,anselmann,cleveland,abdurashitov},
atmospherically-produced  neutrinos \cite{fukuda2,hirata,becker}, and
accelerator-generated neutrinos \cite{athanassopoulos}. These observations
raise the question whether they can all be explained  by a simple
three-neutrino oscillation model \cite{teshima,fogli,acker,cardall,conforto,harrison}. So far,
data on  the atmospheric neutrino anomaly have been fitted only to a 
two-neutrino oscillation hypothesis. We believe that this may lead  to an
incorrect conclusion regarding the oscillation process and  its associated
parameters. 
 
Several aspects of the Super-Kamiokande data motivate us to propose  that
muon-neutrino to tau-neutrino oscillations are observed with  $\Delta m^2$
around 0.4 $eV^2$ and that, furthermore, oscillations  between muon  neutrinos
and electron neutrinos (and vice-versa) are seen with $0.0001 <\Delta m^2
<0.001~eV^2$. First, low-energy ($<E>$=0.75 GeV) ``overhead"  (L$\sim$20 km)
events show a highly significant deficit of muons, relative to electrons,  indicating  an oscillation
process with $\Delta m^2>$0.04 $eV^2$. This would be  consistent with the positive
oscillation result obtained in the LSND experiment  which, when combined with
negative results from other experiments,  requires $\Delta m^2\sim0.4~eV^2$.
Second, Super-Kamiokande observes an additional, significant muon deficit for
``upward" events for which the pathlength L$\sim$10,000 km. These events also
show some enhancement of electron events, suggestive of muon-neutrino to
electron-neutrino oscillations. 

\section{Assumed Model for Neutrino Oscillations}

          In what follows, we assume the simplest general case of 
       three-neutrino mixing:

\begin{equation}
\left(\begin{array}{c}
v_e \\
v_\mu\\
v_\tau\\\end{array}\right) =
\left(\begin{array}{ccc}
U_{11} &U_{12} &U_{13}\\
U_{21} &U_{22} &U_{23}\\
U_{31} &U_{32} &U_{33}\\\end{array}\right)~~
\left(\begin{array}{c}
v_1\\
v_2\\
v_3\\\end{array}\right)
\end{equation}

\noindent
where $v_e, v_\mu, v_\tau$ are, respectively, the electron, muon, and tau
neutrino states, and $v_1, v_2,$ and $v_3$ are the neutrino mass  eigenstates
with masses $m_1<m_2<m_3$. For simplicity, the unitary  mixing matrix U is
assumed to be real and expressed as:

\begin{equation}
U = \left(\begin{array}{ccc}
c_1 c_3 &~~s_1 c_3 &~~s_3\\
-s_1 c_2-c_1 s_2 s_3 &~~c_1 c_2-s_1 s_2 s_3 &~~s_2 c_3\\
s_1 s_2-c_1 c_2 s_3 &~~-c_1 s_2-s_1 c_2 s_3 &~~c_2 c_3\\\end{array}\right)
\end{equation}

\noindent
where c stands for cosine and s for sine of the appropriate 
       three angles, $\theta_1, \theta_2$, and $\theta_3$.

          Oscillations of neutrinos of type $\alpha$ into neutrinos of type 
$\beta$ occur with a probability given by:

\begin{equation}
P(\alpha\rightarrow \beta) = \delta_{\alpha\beta} -4
{\lower8pt\hbox{$\sum\limits^3_{i=1} \sum\limits^3_{j=1}\atop i>j$}} U_{\alpha
i} U_{\beta i}  U_{\alpha j} U_{\beta j} \sin^2 \biggl[{(m^2_i - m^2_j)L\over
4E}\biggr]
\end{equation}

\noindent
where $\delta_{\alpha\beta}$ = 1 if $\alpha =\beta$ and $\delta_{\alpha\beta} =
0$ if $\alpha\neq\beta$, L is the length of the neutrino flight path, and E is the  neutrino
energy. If m is given in eV, E in GeV, and L in km, then  the factor of L/4E in
the sine term becomes 1.27 L/E.

We now discuss the interpretation of various observations in terms of this
simple model. Given our hypothesis regarding the atmospheric neutrino data, we
shall assume that $m_3\gg m_2>m_1$.  We define:

\begin{eqnarray}
m^2 &\equiv m^2_2-m^2_1 \nonumber \\
M^2 &\equiv m^2_3-m^2_2 \nonumber \\
{\rm~and~}M^2+m^2 &= m^2_3-m^2_1
\end{eqnarray}

with $M^2\gg m^2$. 

For the mixing parameters $U_{ij}$ obtained in our analysis, the value of $M^2$
allowed by LSND is highly constrained by the results \cite{achkar,dydak} of the
Bugey reactor experiment $(M^2>0.2~eV^2)$ and of the CDHS  oscillation search
$(M^2<0.4~eV^2)$. We shall assume that $M^2=0.4~eV^2$ for reasons noted below.
To produce a significant effect in Super-Kamiokande the value of $m^2$  must be
at the 0.0001 $eV^2$ level or larger. The CHOOZ reactor experiment
\cite{apollonio} has shown that $\Delta m^2$ is smaller than 0.001 $eV^2$.  We
shall therefore require that 0.0001$<m^2<0.001~eV^2$.

\section{LSND}

The LSND probability to observe oscillations of muon neutrinos into electron
neutrinos is given by:


\begin{eqnarray}
P(LSND) = &-4 U_{23} U_{13} U_{21} U_{11} \sin^2 \biggl[{(M^2+m^2) 
L\over 4E}\biggr] \nonumber \\
 &-4 U_{23} U_{13} U_{22} U_{12} \sin^2 \biggl({M^2 L\over 
4E}\biggr) \nonumber \\
 &-4 U_{22} U_{12} U_{21} U_{11} \sin^2 \biggl({m^2 L\over 4E}\biggr)
\end{eqnarray}

In the LSND experiment \cite{athanassopoulos} L=30 m and $36<E<60$ MeV, so that
for $m^2<0.001~eV^2$ the third term in the above expression can be neglected. 
The first two terms can be combined by approximating $M^2+m^2 \approx M^2$ and
using the unitarity of the mixing matrix:

\begin{equation}
P(LSND) = 4 [U_{23} U_{13}]^2 \sin^2 \biggl({M^2 L\over
4E}\biggr) = 4 (s_2 c_3 s_3)^2 \sin^2 \biggl({M^2 L\over 4E}\biggr)
\end{equation}

The measured value is P(LSND)=0.0031 $\pm$0.0011(stat)  $\pm$0.0005(syst). 
When other neutrino data are also considered, the smallness of P(LSND)  arises
in part from the smallness of $s_3$. For $M^2=0.4~eV^2$, L=30 m,  and a mean
value of E=42 MeV, we obtain $s_2 c_3 s_3$=0.0784.

\section{Atmospheric Neutrinos}

Super-Kamiokande will observe oscillations of muon neutrinos into tau neutrinos
with a probability given by:

\begin{eqnarray}
P(SK; v_\mu \rightarrow v_\tau) = &-4 U_{23} U_{33} U_{21} U_{31} \sin^2 
\biggl[{(M^2 + m^2)L\over 4E}\biggr] \nonumber \\ 
&-4 U_{23} U_{33} U_{22}
U_{32} \sin^2 \biggl({M^2 L\over 4E}\biggr) \nonumber \\ 
&-4U_{22} U_{32}
U_{21} U_{31} \sin^2 \biggl({m^2 L\over 4E}\biggr)
\end{eqnarray}

\noindent
For the oscillation parameters considered in this paper, the third term can be
neglected and the first two terms combined to obtain:

\begin{equation}
P(SK; v_\mu \rightarrow v_\tau) = 4[s_2 c_2 c^2_3]^2 \sin^2 {M^2L\over 4E} 
= 2[s_2 c_2 c^2_3]^2
\end{equation}

\noindent
where the variation in E and L allow us to average the sine-squared factor to a
value of 0.5.

The smallness of the LSND effect and the failure to observe electron-neutrino
disappearance in reactor experiments \cite{achkar,apollonio} indicate that the
overall deficit of muons observed in the Super-Kamiokande atmospheric data for
zenith angles satisfying $\cos\theta_z>-0.6$ results primarily from 
oscillations of muon neutrinos into tau neutrinos.  Using the result that 
P(SK)=0.30 for such events \cite{fukuda2}, we obtain $s_2 c_2 c^2_3=0.387$.  
Combining this with the LSND results that $s_2 c_3 s_3=0.0784$, we find
$\theta_2=26.5^\circ$ and $\theta_3=10.3^\circ$:

We quote results using central values for various parameters.  Clearly, the
uncertainty in an input such as P(LSND) will  produce uncertainties in a
parameter such $\theta_3$ (of order $\pm2^\circ$).  Our purpose is to show that
these central values can account for essentially all observations.

Super-Kamiokande will also observe oscillations between muon neutrinos and
electron neutrinos with a probability given by:

\begin{eqnarray}
P(SK; v_\mu \leftrightarrow v_e) = &-4 U_{23} U_{13} U_{21} U_{11} \sin^2 
\biggl[{(M^2 + m^2)L\over 4E}\biggr] \nonumber \\
&-4 U_{23} U_{13} U_{22} U_{12} \sin^2 \biggl({M^2 L\over 4E}\biggr) 
\nonumber \\
&-4 U_{22} U_{12} U_{21} U_{11} \sin^2 \biggl({m^2 L\over 4E} \biggr)
\end{eqnarray}

The sine-squared factors of the first two terms of this expression each average
to a value of 0.5 and the two terms can be combined to yield:

\begin{equation}
P(SK; v_\mu \leftrightarrow v_e) = 2[U_{23} U_{13}]^2 - 4 U_{22} U_{12} 
U_{21} U_{11} \sin^2 \biggl({m^2 L\over 4E}\biggr)
\end{equation}

For our values of $\theta_2$ and $\theta_3$, the first term is small
($\sim$0.02), justifying our assumption that, for most of the zenith-angle
range, the atmospheric muon anomaly is caused by muon-neutrino to tau-neutrino
oscillations.  However, the second term will produce observable effects for
``upward'' going events for which ${m^2L\over 4E}$ is assumed to be of order
one or greater.  We return to this point in the discussion below.

We have examined the question of matter (MSW) effects on oscillations
between atmospheric electrons and muon neutrinos \cite{barger}. For the
large mixing angles obtained for our model, resonance effects do not
have a significant impact on our analysis.  However, for $m^2$ values
near 0.0001 $eV^2$, oscillations between electron and muon neutrinos
are strongly damped for neutrino energies above 1 GeV. Such damping
effects decrease rapidly with increasing values of $m^2$, and are
insignificant at the upper end (0.001 $eV^2$) of the $m^2$ range
considered here.

\section{Solar Neutrinos}

The MSW mechanism is not expected to have a significant impact on
solar neutrinos for the large mass-squared
differences considered in this paper, so that the probability for a solar
electron neutrino to remain as such is:

\begin{eqnarray}
P(solar; v_e \rightarrow v_e) = &1-4 [U_{13} U_{11}]^2 \sin^2
\biggl[{(M^2+m^2) L\over 4E}\biggr] \nonumber \\
&-4 [U_{13} U_{12}]^2 \sin^2 \biggl({M^2 L\over 4E}\biggr) \nonumber \\
&-4 [U_{12} U_{11}]^2 \sin^2 \biggl({m^2 L\over 4E}\biggr)
\end{eqnarray}

\noindent
For the E and L of solar neutrinos detected on Earth, the sine-squared
       terms in the above expression average each to 0.5 and:

\begin{equation}
P(solar; v_e \rightarrow v_e) = 1-2 (s_3 c_3)^2 - 2(s_1 c_1 c^2_3)^2
\end{equation}
       
Within the uncertainties of solar neutrino theory, essentially  all solar
neutrino observations (with the possible exception of those  from the Homestake
experiment) are consistent with P(solar)=0.5 \cite{98conf}.  Using this value
and our previous result for angles $\theta_2$ and $\theta_3$, we obtain
$\theta_1=37.6^\circ$.

\section{Reactor Experiments}

The general expression derived for solar neutrinos also applies to  reactor
experiments. Such experiments have not been sensitive to  $m^2$
values less than 0.001 $eV^2$. However, oscillations with  $M^2=0.4~eV^2$ are,
in principle, observable. For values of E and L that allow an averaging of the
oscillation factor, we expect:

\begin{equation}
P(reactor; v_e \rightarrow v_e) = 1-2 (s_3 c_3)^2
\end{equation}

With $\theta_3 = 10.3^\circ$, we predict P(reactor) = 0.94 compared to measured
values of 0.99$\pm$0.01(stat)$\pm$0.05(syst) for Bugey \cite{achkar} and 
0.98 $\pm$0.04(stat) $\pm$0.04(syst) for CHOOZ \cite{apollonio}. Given 
the size of the errors, this is reasonable  if not
perfect agreement.  In the Bugey experiment, the detectors were sufficiently
close to the reactor to be sensitive, in principle, to modulations in the
positron spectrum produced by the $\sin^2 {M^2L\over 4E}$ factor.  We discuss
this below.

\section{Summary and Discussion}

We have found that most features of oscillation-related neutrino data can be
explained by the following mixing matrix, corresponding to the angles
$\theta_1=37.6^\circ, \theta_2=26.5^\circ$, and $\theta_3=10.3^\circ$.

\begin{equation}
U = \left(\begin{array}{ccc}
0.78 &~~0.60 &~~0.18\\
-0.61 &~~0.66 &~~0.44\\
0.15 &~~-0.45 &~~0.88\\ \end{array}\right) \\
\end{equation}

\noindent
and by assuming that neutrino masses $m_1, m_2, m_3$ satisfy 
$m_3>>m_2>m_1$, that $0.0001<(m^2_2-m^2_1)<0.001~eV^2$, and that
$M^2=m^2_3-m^2_2 =0.4~eV^2$. This choice of parameters appears to be in
reasonable  accord with essentially all neutrino observations, including the
zenith-angle behavior of the Super-Kamiokande atmospheric neutrino data for
``upward" going ($\cos\theta_z <-0.6$) events.  We demonstrate this by assuming
that L is sufficiently  large that all sine-squared factors involving
mass-squared differences average to 0.5.  We find that:

\begin{eqnarray}
P_{\mu\mu} &=& P(SK; v_\mu \rightarrow v_\mu) = U^4_{21} + U^4_{22} +
U^4_{23} \nonumber \\
P_{ee} &=& P(SK; v_e \rightarrow v_e) = U^4_{11} + U^4_{12} + 
U^4_{13} \nonumber \\
P_{e\mu} &=& P_{\mu e} = P(SK; v_\mu \leftrightarrow v_e) = 2 [U_{23} U_{13}]^2 
-2 U_{22} U_{12} U_{21} U_{11}
\end{eqnarray}

\noindent
The double ratio R for these ``upward" going events:

\begin{equation}
R={(N_\mu/N_e) {\rm~measured}\over (N_\mu/N_e) {\rm ~no~oscillation}}
\end{equation}

\noindent
can be estimated to be:

\begin{equation}
R={(2P_{\mu\mu} + P_{e\mu})/(P_{ee}+2P_{\mu e})\over 2}
\end{equation}

\noindent
given the fact that atmospheric $v_\mu$ are produced at approximately twice
the rate of $v_e$.  We obtain the result that R=0.44 compared to the measured value of
0.41$\pm$0.04 at the largest zenith angles.

We show in Fig. 1 the predicted ratio R as a function of L/E for three
values of $m^2 \equiv m^2_2-m^2_1$.  The curve of R vs.~L/E has been smeared
by a resolution function that is Gaussian in ln(L/E) with a width that
corresponds to a factor of three uncertainty in the L/E measured for any given
event.  In an experiment such as Super-Kamiokande the angle between the
incoming neutrino and the direction of the detected lepton can be quite large
\cite{fukuda2}, leading to a significant uncertainty in L.  The general shape of
the curve of R vs.~L/E in Fig.~1 depends only weakly on the assumed smearing,
with factors of two or four giving qualitatively the same results as our
assumed factor of three.  We have also ploted in Fig. 1 the latest
Super-Kamiokande results \cite{fukuda2} for which a reasonably good fit
is obtained for $m^2$ values around 0.0003 $eV^2$.

\begin{figure}[t]
\epsfxsize=6.0in
\begin{center}
\includegraphics[scale=0.75]{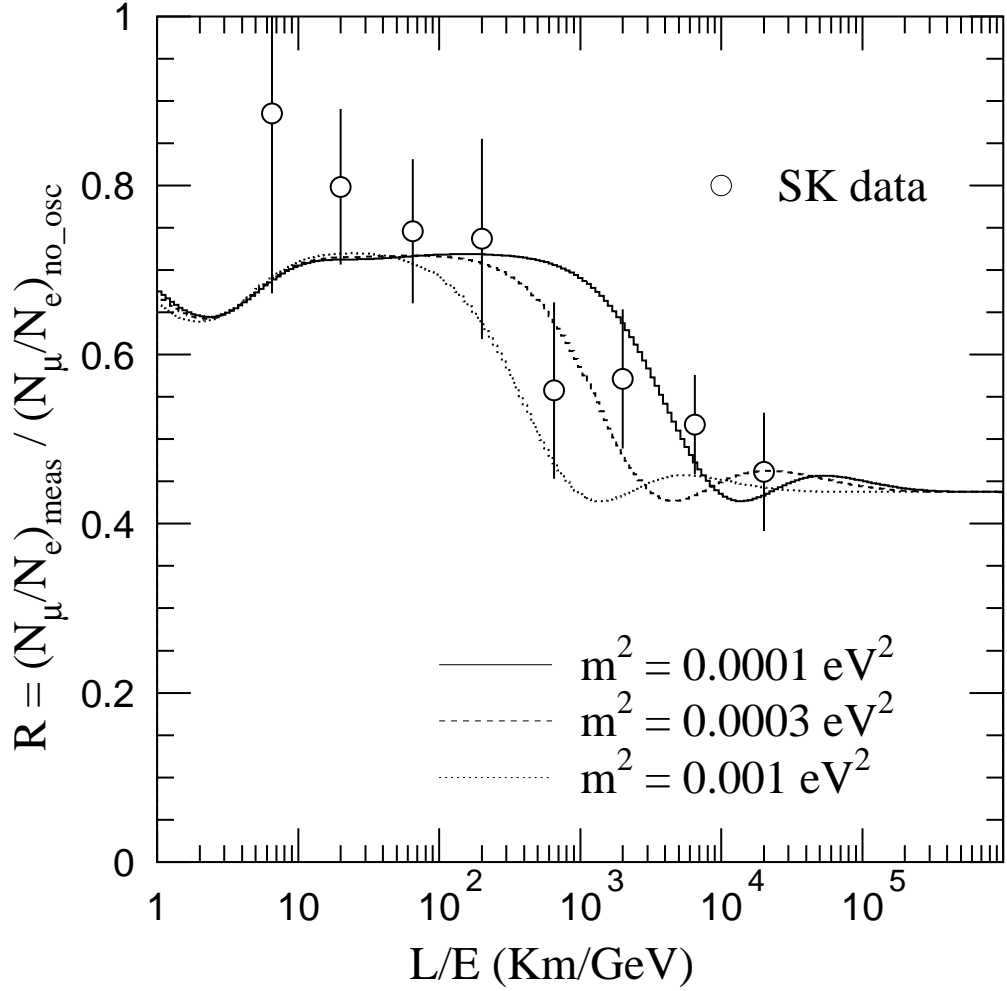}
\end{center}
  \vspace{-0.5cm}
\caption{The predicted double ratio R as a function of L/E for the mixing matrix with
$\theta_1=37.6^\circ, \theta_2=26.5^\circ, \theta_3=10.3^\circ$ and for three
values of $m^2 \equiv m^2_2-m^2_1$. The value of $M^2\equiv
m^2_3-m^2_2 = 0.4\ eV^2$. The data are the most recent
Super-Kamiokande results~\cite{fukuda2}.}
\medskip
\end{figure}

   The most recent Super-Kamiokande paper \cite{fukuda2} makes available the ratio
of the number of measured to the number of expected events, assuming no
oscillations, for electrons and muons separately. We show these data in
Fig.~2, with the overall flux normalization reduced by 12\% to give a best
fit to the three-neutrino mixing model presented in this paper. Such a
normalization shift is well within the errors of the theoretical flux
assumed by the Super-Kamiokande Collaboration. Our model does not give a 
very good fit to the individual electron and muon ratios. However, 
it is important to note that the double ratio R shown in Fig. 1, for which we 
obtain a good fit, is much less sensitive to systematic errors than are
the individual electron and muon ratios. Until these systematic errors
are reduced by improved measurements of atmospheric cosmic-ray particle 
production, one should keep an open mind about the significance of individual 
distributions in the atmospheric neutrino data.
\begin{figure}[t]
\epsfxsize=6.0in
\begin{center}
\includegraphics[scale=0.75]{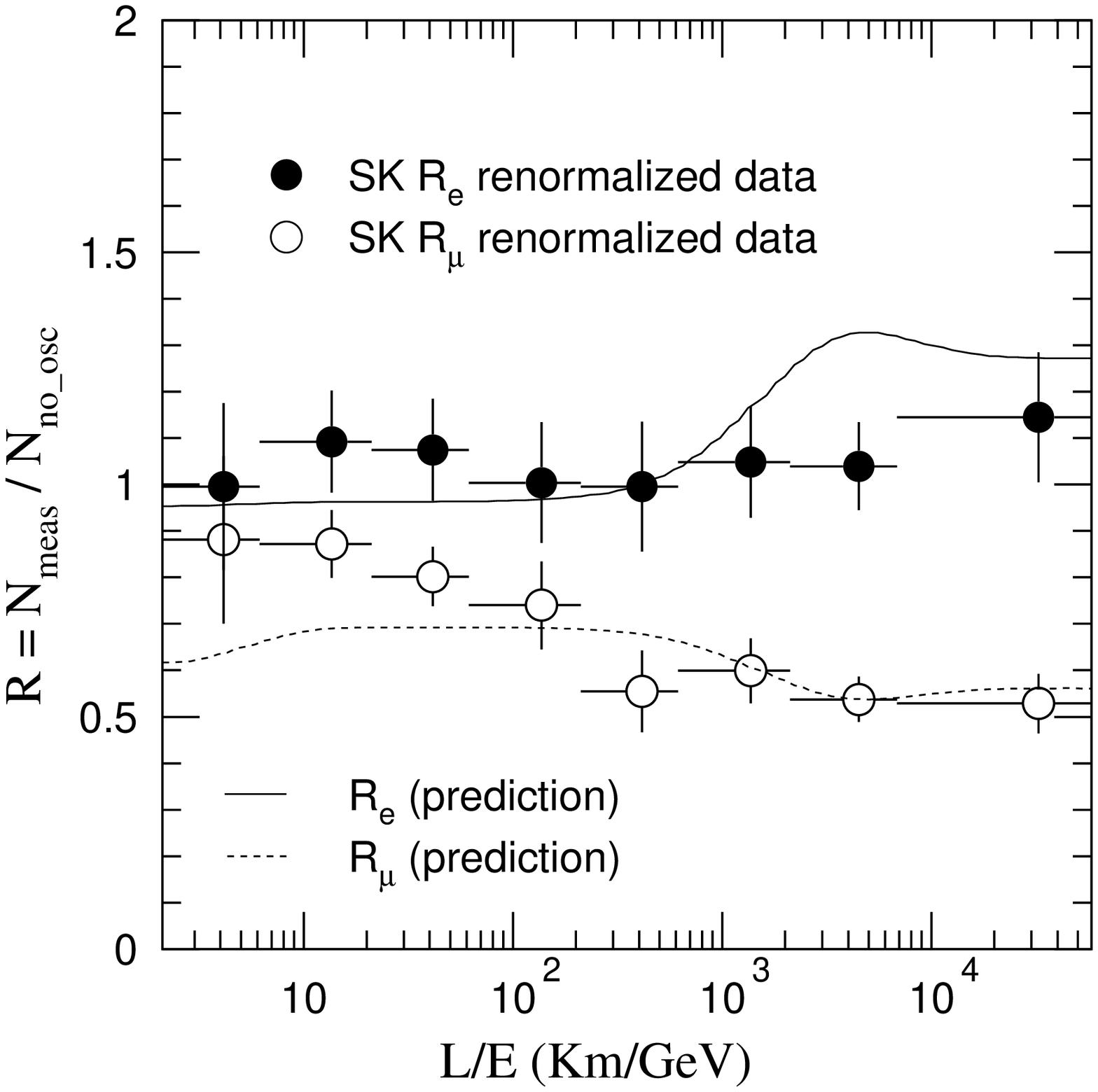}
\end{center}
  \vspace{-0.5cm}
\caption{Separate Super-Kamiokande electron and muon data with overall
flux normalization reduced by 12\% for a best fit to the three-neutrino
mixing model presented in this paper with $m^2 = 0.0003$ eV$^2$.}
\medskip
\end{figure}

As with previous attempts to explain all neutrino anomalies with a simple
three-neutrino mixing model, the one outlined in this paper is subject to the
resolution of various experimental inconsistencies and ambiguities. For
example, we have assumed that the LSND result is an oscillation effect rather
than unexplained background. It must be noted that the KARMEN experiment has
not confirmed the LSND result. For our assumed parameters, recent KARMEN data
\cite{zeitnitz} should have yielded approximately two signal events in addition
to roughly three  background events, whereas no oscillation candidate events
were seen. However, the LSND collaboration has reported additional
evidence for oscillations in events in which the neutrino was produced by
decays in flight~\cite{athan2}.

With respect to the atmospheric neutrino data of Super-Kamiokande, our model
with $\theta_1=37.6^\circ, \theta_2=26.5^\circ, \theta_3=10.3^\circ$,
$M^2\equiv m^2_3-m^2_2 = 0.4~eV^2$ and
$m^2\equiv m^2_2-m^2_1 =0.0003~eV^2$ gives a good overall fit to the data.  To
distinguish our model from one with a single oscillation process with a small
$\Delta m^2 (\sim~0.0022~eV^2)$ requires precise measurements of multi-GeV,
``overhead" ($\cos \theta_z \sim 1$) events.  Our model predicts R$\sim$0.72
for such events whereas the small-$\Delta m^2$ model yields R$\sim$1.
Present Super-Kamiokande data are inconclusive in this low-L/E region.

Our model requires that solar neutrino deficits observed on Earth should show
no energy dependence and no MSW effects. The data from the gallium experiments
(GALLEX and SAGE) and from Super-Kamiokande, which sample very different parts
of the solar neutrino energy spectrum, are in general agreement with this
prediction. However, the Homestake Cl-37 experiment has historically reported a
substantially larger solar neutrino deficit \cite{cleveland2} than is observed
in these other experiments.  We note, however, that the measured Homestake
solar neutrino flux of 2.56 $\pm$0.16(stat) $\pm$0.16(syst) SNU represents
0.403 $\pm$ 0.025 $\pm$0.025 of the expected flux computed by Turck-Chieze and
Lopes \cite{turck}, not far from the value 0.50 assumed in this paper.   A
final interpretation of all the data awaits a better understanding of both the
theory and of possible systematic errors in various experimental results.

Our model agrees within one standard deviation with the fluxes measured in
reactor experiments.  However, for $\theta_3=10.3^\circ$, the Bugey results
\cite{achkar} would be expected to show some modulation (at the level of several
percent) of the shape of the measured positron spectrum.  We note, however, that
statistical and systematic uncertainties in the LSND result permit values of
$\theta_3$ as small as $8^\circ$, which would reduce such modulations below the
level of detectability.  Our choice of $M^2=0.4~eV^2$, the maximum allowed by
the CDHS oscillation search \cite{dydak}, is dictated by the requirement that
$\theta_3$ be small.

We note that our model predicts large effects for  terrestrial,
long-baseline experiments such as K2K, MINOS, and those planned for Gran Sasso.
Furthermore, if muon neutrinos oscillate into tau neutrinos with $M^2=0.4~
eV^2$, then observable effects are just beyond the reach of the present
short-baseline experiments CHORUS and NOMAD but accessible to a new, improved 
experiment such as proposed by the TOSCA collaboration.

Finally, we remark that our model differs from previous versions of
simple three-neutrino
mixing \cite{teshima,fogli,acker,cardall,conforto,harrison} in its
unique combination of a $M^2$ value that can explain the LSND results
and a $m^2$ value significantly larger than is usually assumed in
solar neutrino models.   It is, of course, possible that eventually no
simple three-neutrino mixing model will give a satisfactory account of
past, present and future experimental results.  In that case, more
complicated models will have to be considered \cite{Geiser}.

\section{Acknowledgement}
We thank S.B.~Kim for helpful discussions.  This work was performed under the
sponsorship of the U.S.~Department of Energy, contract DoE FG02-95ER40899.



\begin{thebibliography}{99}
\bibitem{98conf} At the Neutrino '98 Conference (Takayama, Japan) the 
following results
were reported for solar neutrino results, where SSM stands for standard solar
model prediction:
\begin{description}
\item{~~~~~} Super-Kamiokande:  data/SSM = 0.474
\item{~~~~~} GALLEX:  data/SSM = 77.5/132 = 0.587
\item{~~~~~} SAGE:  data/SSM = 66.6/137 = 0.486
\item{~~~~~~}with overall errors on data/SSM in the range of 10\% to 15\%. 
\end{description}
\end{thebibliography}
\end{document}